\documentclass[aps,12pt,preprintnumbers,showpacs]{revtex4}
\usepackage{amsmath,amssymb,bm,epsfig}
\renewcommand\d{\partial}
\newcommand\p{{\vec p}}
\newcommand\q{{\vec q}}

\newcommand\x{\bm{x}}
\newcommand\vgamma{{\vec\gamma}}

\newcommand{\diracsl}[1]{#1\hskip-7pt \diagup}

\newcommand\Tr{\textrm{Tr}\,}
\newcommand\ksl{{k\hskip-9pt\diagup}}
\newcommand\qsl{{\diracsl q}}
\newcommand\psl{{\diracsl p}}
\newcommand\stru{\rule[-15pt]{0pt}{15pt}}

\begin{document}
\preprint{INT-PUB 07-07}
%\title{A New Type of Quantum Critical Behavior\\ in Graphene-Like Systems}
%\title{Graphene as a system near quantum criticality}
\title{Quantum critical point in graphene approached in the limit of
infinitely strong Coulomb interaction}
\author{D.~T.~Son}
\affiliation{Institute for Nuclear Theory, University of Washington,
Seattle, Washington 98195, USA}
\date{January 2007}
\begin{abstract}
\noindent
Motivated by the physics of graphene, we consider a model of $N$
species of $2{+}1$ dimensional four-component massless Dirac fermions
interacting through a three dimensional instantaneous Coulomb interaction.  
We show
that in the limit of infinitely strong Coulomb interaction, the system
approaches a quantum critical point, at least for sufficiently large
fermion degeneracy.  In this regime, the system exhibits invariance
under scale transformations in which time and space scale by different
factors.  The elementary excitations are fermions with dispersion
relation $\omega\sim p^z$, where the dynamic critical exponent $z$
depends on $N$.  In the limit of large $N$, we find
$z=1-4/(\pi^2N)+O(N^{-2})$.  We argue that due to the numerically
large Coulomb coupling, graphene (freely suspended) in vacuum stays
near the scale-invariant regime in a large momentum window, before
eventually flowing to the trivial fixed point at very low momentum
scales.

\end{abstract}
\maketitle

\section{Introduction.}

The physics of graphene, which has recently been realized in
experiment~\cite{Novoselov-PNAS}, has currently attracted
considerable interest.  A graphene sheet is a two dimensional (2D) hexagonal
lattice of carbon atoms, and its defining feature is the existence of
two special points in the Brillouin zone around which the electron
energy, according to band-structure calculations, should have a linear
dependence on its momentum (``Dirac cones''), as in relativistic
theories~\cite{Semenoff:1984dq}.  Many interesting behaviors, for
example, the quantum Hall effect at half-integer filling fractions, are
attributed to the quasirelativistic behavior of the low energy
excitations~\cite{Novoselov-Nature,Kim-Nature,Gusynin:2005pk}.

Graphene, on the other hand, differs from relativistically invariant
systems in one crucial aspect.  Namely, the electromagnetic
interaction between fermion quasiparticles is mediated by the photon
whose velocity $c$ is practically infinite compared to the fermion
velocity $v\approx c/300$.  The interaction therefore is an
instantaneous Coulomb repulsion which breaks relativistic invariance.

As we shall see later, the importance of the Coulomb interaction is
controlled by the parameter
\begin{equation}
  \lambda = \frac{e^2N}{16\epsilon_0\hbar v}\,,
\end{equation}
where $N=2$ is the spin degeneracy.  It is similar to the fine
structure constant, but the speed of light has been replaced by the
Fermi velocity $v$.  As the result, for graphene in vacuum this
parameter is around 3 or 4 (it is reduced if graphene resides on a
substrate with a large dielectric constant).  The largeness of
$\lambda$ makes unreliable any calculation based on the simple
perturbation theory in the interaction strength, and seems to indicate
that real graphene is hopelessly beyond quantitative theoretical
control.

In this paper, we show that in the idealized limit $\lambda\to\infty$
the system becomes, in a certain sense, simple again.  We shall argue
that, at least for sufficiently large $N$, and perhaps even for $N=2$,
$\lambda\to\infty$ is a limit where the system is tuned to quantum
criticality.

Quantum critical points~\cite{Sachdev} play an important role in
condensed matter physics.  In many cases, they are described by
relativistically invariant conformal field theories.  The relativistic
invariance of these theories is not connected to the relativity of
space-time, but is a manifestation of an emergent Lorentz symmetry.
% of
%the infrared physics, where there is a special velocity $v$ which
%plays the role of the effective speed of light.  
In our case, the
infinitely strong Coulomb interaction destroys the relativistic
invariance.  Instead, we shall see that the quantum critical point
$\lambda=\infty$ is characterized by a nontrivial dynamic critical
exponent $z$, whose value is computable in the large-$N$ limit,
\begin{equation}
  z = 1 - \frac4{\pi^2N} + O(N^{-2}).
\end{equation}
The elementary excitations are fermions with dispersion relation
$\omega\sim p^{z}$, instead of the linear dispersion $\omega=vp$.  The
Dirac cones are therefore replaced by ``Dirac cusps.''  Moreover,
the full dynamics is invariant under the scale transformation
\begin{equation}\label{scale-z}
 t \to l^z t, \qquad \x \to l x
\end{equation}
instead of the usual ``relativistic'' scale transformation
\begin{equation}\label{scale-1}
 t \to l t, \qquad \x \to l x.
\end{equation}

The limit $\lambda\to\infty$ can therefore be considered as a nice
idealization of graphene.  While it is only an idealization, we think
that its simplicity justifies our study.  Moreover, one may hope that
once the limit $\lambda\to\infty$ is understood, the case of finite
but large $\lambda$ can be accommodated by treating $\lambda^{-1}$ as
a small coefficient of a relevant deformation that takes the system
away from the critical point.

The structure of this paper
is as follows.  In Sec.~\ref{sec:model}, we describe our model.  We
write down the renormalization group (RG) equation in
Sec.~\ref{sec:RG} and discuss the strong-coupling fixed point where 
the scaling behavior~(\ref{scale-z}) is realized.  The running of the
fermion velocity at finite Coulomb coupling is considered in 
Sec.~\ref{sec:finite-lambda}.
In Sec.~\ref{sec:graphene}, we discuss
finite $N$ and the relevance of our model to real graphene.  We
conclude in Sec.~\ref{sec:conclusion}.  While a large fraction of
technical calculations presented in this paper are not new, we
believe that the strong-coupling limit has not been previously
identified as a quantum critical point.

\section{Model}
\label{sec:model}

We shall consider a system of $N$ (2+1) dimensional [(2+1)D] 
four-component massless Dirac
fermions with velocity $v$ interacting through an instantaneous three
dimensional (3D)
Coulomb interaction.  The Euclidean action is (in this paper, we set
$\hbar=1$)
\begin{equation}\label{S}
  S = -\sum_{a=1}^N\int\!dt\,d^2x\, (\bar\psi_a\gamma^0\d_0\psi_a +
        v\bar\psi_a\gamma^i\d_i\psi_a  +iA_0 \bar\psi_a \gamma^0\psi_a)
  + \frac1{2g^2}\int\!dt\,d^3x\, (\d_i A_0)^2.
\end{equation}
Our notations are as follows.  The fields $\psi_a$ are four-component
fermion fields, and $a$ labels different species of fermions.  In real
graphene $N$, is equal to 2 due to the spin degeneracy.
The $\gamma$'s are (2+1)D Dirac matrices satisfying
$\{\gamma^\mu,\gamma^\nu\}=2\delta^{\mu\nu}$, and can be chosen as
$4\times4$ matrices, e.g., $\gamma^0=\sigma^3\otimes\sigma^3$,
$\gamma^i=\sigma^i\otimes\openone$.  
Each fermion can be thought of as a pair of two-component
fermions with opposite parities~\cite{Nielsen:1980rz,Nielsen:1981xu}.
$A_0$ is the Coulomb potential.  The action~(\ref{S}) contains a (2+1)D 
part, which contains the kinetic term for the fermion and the 
interaction between the fermion and the Coulomb potential, and a (3+1)
dimensional
part, which is the kinetic term for the Coulomb potential.
A more general model was considered in Ref.~\cite{YeSachdev} in the
context of the quantum Hall effect.  
%Note that the fermions propagate
%in two spatial dimensions, but the Coulomb potential $A_0$ propagates
%in all three.  
If graphene is in vacuum (as it is the case for freely
suspended graphene sheets~\cite{suspended}), then $g^2=e^2/\epsilon_0$,
where $e$ is the electron charge and $\epsilon_0$ is the vacuum
permeability.  In the presence of a substrate with a dielectric
constant $\varepsilon$, the effective charge is reduced,
\begin{equation}\label{g}
  g^2 = \frac2{1+\varepsilon}\frac{e^2}{\epsilon_0}\,.
\end{equation}
In Eq.~(\ref{S}), we have not included any contact four-Fermi
interactions, which are irrelevant at weak coupling.  At strong
coupling, however, these interactions develop nontrivial fixed
points~\cite{Herbut:2006cs}.  We shall assume that these four-Fermi
interactions start out with small couplings and flow to the trivial
fixed point.

We will be particularly interested in the limit of infinite Coulomb
repulsion $g^2\to\infty$.  In this limit, there is no kinetic term for
$A_0$ in the bare action~(\ref{S}); however, an effective kinetic term
will be generated by the fermion loop.  We find it useful to keep
$g^2$ large but finite for the purpose of regularization and for the
discussion of the real graphene.

Without the coupling to the scalar potential $A_0$, the theory is that
of free Dirac fermions which is invariant under the relativistic scale
transformation~(\ref{scale-1}).  Our goal is to see that after coupling
to the scalar potential the system remains scale invariant, but under
a more general type of scale transformation with a (generally)
fractional $z$.

%It it quite common that 
%That the operation of coupling a scale-invariant theory to a field with no
%bare kinetic term, one can obtain a new scale-invariant theory has been 
%seen in many occasions.  
A relativistic counterpart of our model was considered previously.  In
Refs.~\cite{Templeton:1981ei,Templeton:1981yp} free fermions are
coupled to a gauge field $A_\mu$.  At sufficiently large $N$, the infrared
limit of the new system is described by a conformal field theory.
(For $N$ smaller than some critical value, which is still not exactly
known, it was argued that the system develops a mass
gap~\cite{Appelquist:1988sr}.)
%This operation can be identified as
%the ${\cal S}$ operator of an $SL(2,Z)$ group of mappings between
%(2+1)D conformal field theories~\cite{Witten:2003ya}.  The system
%describing fermions at Feshbach resonance can be obtained by coupling
%free fermions to a complex scalar field in the Cooper
%channel~\cite{Nishida:2006br,Nishida:2006eu}.  In dimensional
%regularization, the limit when the scalar field has no bare kinetic
%term corresponds to the infinite scattering length between fermions in
%the resulting theory.
The difference between our case and the case considered in
Refs.~\cite{Templeton:1981ei,Templeton:1981yp} is the lack of Lorentz
invariance due to the instantaneous Coulomb interaction.  The fact that
in graphene Coulomb interaction induces a logarithmic renormalization
of the fermion velocity and leads to logarithmic corrections in
thermodynamics is well known~\cite{Gonzalez:1993uz}.  To make the
paper self-contained, we repeat some of the calculations in the
literature.  To start, let us state the Feynman rules that follow
from Eq.~(\ref{S}),
\begin{itemize}
\item[(i)] The fermion propagator is
\begin{equation}
  G_0(p) = \frac i{\psl} = \frac{i\psl}{p^2}\,.
\end{equation}
Here we use quasirelativistic notation, where $p$ stays for $(p_0,\p)$
(for example $d^3p\equiv dp_0\,d\p$)
and $\psl\equiv\gamma^0\p_0 +v \vgamma\cdot\p$, $p^2\equiv
p_0^2+v^2|\p|^2$, and $\p$ is the 2D momentum vector.
\item[(ii)] The $A_0$ propagator is the integral of the 3D
  propagator $g^2/(p_z^2+|\p|^2)$ over the momentum component
  perpendicular to the 2D plane, $p_z$,
\begin{equation}
   D_0(p) = g^2\int\!\frac{dp_z}{2\pi}\, \frac 1{p_z^2+|\p|^2}
  = \frac {g^2}{2|\p|}\,.
\end{equation}
and is simply the 2D Fourier transform of the function $g^2/(4\pi r)$. 
\item[(iii)] The interaction vertex is $i\gamma^0$.
\end{itemize}
For the simplicity of the notations, one can perform calculations
using the unit system where $v=1$, then restore $v$ at the end results
by dimensionality.  To have analytic control, we shall work in the
large-$N$ limit and then extrapolate to $N=2$. Each fermion loop comes
with a factor of $N$, so in the large-$N$ limit one has to resum the
fermion loop in the photon propagator (this is identical to the random
phase approximation).  The fermion loop is naively divergent; however,
in any gauge-invariant regularization scheme (e.g., dimensional
regularization) it is convergent.  The resummed Coulomb propagator is
(see the Appendix)
\begin{equation}
  D(q) =  \left( \frac{2|\q|}{g^2} + \frac N8 
  \frac{|\q|^2}{\sqrt{q^2}} \right)^{-1}.
\end{equation}
Since $D(q)\sim1/N$, even in the $g^2\to\infty$ limit the interaction
between fermions remains weak at large $N$, enabling a perturbative
calculation.

\section{Renormalization group and the strong-coupling fixed point}
\label{sec:RG}

We shall now perform Wilson RG in our theory at the leading nontrivial
order in $1/N$.  We assume that the theory has a cutoff $\Lambda_0$, and
proceed to integrate out all momenta between $\Lambda_1$ and
$\Lambda_0$, where $\Lambda_1$ is smaller than $\Lambda_0$ by an
exponential factor.  The leading $1/N$ correction to the fermion
kinetic term comes from the one-loop fermion self-energy graph,
\begin{equation}\label{Sigmap}
  \Sigma(p) = -g^2 \int\! \frac{d^3q}{(2\pi)^3}\,
  \frac{\gamma^0(\psl-\qsl)\gamma^0}{(p-q)^2}\,
  \left( 2|\q| + \frac{g^2N}8 \frac{|\q|^2}{\sqrt{q^2}} \right)^{-1},
\end{equation}
where integration is over $q$ in the momentum shell
$\Lambda_1<q<\Lambda_0$.  In RG, we are interested only in the
contribution proportional to $\ln(\Lambda_0/\Lambda_1)$.  

Although the integral can be computed in closed analytic form for any
$\lambda$ (see below), it is instructive to consider first the limit
$\lambda=\infty$.  In this limit, we find
\begin{equation}
  \Sigma(p) = \Sigma_0 \gamma_0 p_0  + \Sigma_1 \vgamma\cdot\p,
\end{equation}
where $\Sigma_0$ and $\Sigma_1$ are represented as integrals,
\begin{equation}
  \Sigma_0 = \frac8N\int\!\frac{d^3q}{(2\pi)^3}\,
  \frac{q_0^2-\q^2}{(q^2)^{3/2}|\q^2|}\,,\qquad
  \Sigma_1 = \frac8N\int\!\frac{d^3q}{(2\pi)^3}\,
  \frac{q_0^2}{(q^2)^{3/2}|\q^2|}\,.
\end{equation}
The integrands scale as $q^{-3}$; therefore, both integrals contain the
factor $\ln(\Lambda_0/\Lambda_1)$.  These correspond to logarithmic
renormalization of the kinetic terms $\bar\psi\gamma^0\d_0\psi$ and
$\bar\psi\gamma^i\d_i\psi$.  The fact that these terms are renormalized
differently means that the fermion velocity changes under the RG.

However, it is easy to see that $\Sigma_0$ and $\Sigma_1$ contain an
additional logarithmic divergence due to the singularity of the
integrands in the limit $|\q|/q_0\to0$.  This singularity can be
traced to the fact that the Coulomb interaction is unscreened in the
finite frequency, zero wave number limit.  However, in this limit the
only effect of the gauge field is to phase rotate the fermion
operator $\psi$; therefore, the logarithmic singularity associated with
the $|\q|/q_0\to0$ limit should disappear in any gauge-invariant
quantity, for example, in the fermion velocity $v$.  Indeed, the
renormalization of $v$ depends on $\Sigma_1-\Sigma_0$, which is free
from the $\q=0$ singularity:
\begin{equation}
  \Sigma_1 - \Sigma_0 = \frac8N \int\!\frac{d^3q}{(2\pi)^3}\, 
  \frac1{(q^2)^{3/2}} = \frac4{\pi^2N}\ln\frac{\Lambda_0}{\Lambda_1}\,.
\end{equation}
The RG equation for the velocity becomes
\begin{equation}\label{RG-v}
  p\frac{\d v(p)}{\d p} = -\frac4{\pi^2N} v(p),
\end{equation}
which implies that in the limit of infinitely strong Coulomb coupling
$\lambda\to\infty$, the velocity has a finite anomalous dimension
$\gamma_v=-4/(\pi^2N)$.  The solution to the RG equation~(\ref{RG-v})
is
\begin{equation}
  v(p) = \textrm{const}\times p^{-4/(\pi^2 N)}.
\end{equation}
Since the velocity is the slope of the dispersion curve, one concludes
that the fermion dispersion relation has the form
\begin{equation}\label{dispersion}
  \omega = \textrm{const}\times p^z,
\end{equation}
with the dynamic critical exponent $z$ being
\begin{equation}\label{z}
  z = 1+\gamma_v = 1-\frac4{\pi^2 N} + O(N^{-2}).
\end{equation}
Note that the argument of Ref.~\cite{Herbut-when} that requires $z=1$
does not apply to our case, since the fixed point here is at infinite
Coulomb coupling $g\to\infty$.

Some physical consequences of Eq.~(\ref{z}) need to be
mentioned. Since $z<1$, the quasiparticle is stable, since its decay
into two or more other quasiparticles is forbidden by energy and
momentum conservation.  The specific heat has a power-law behavior at
small temperature $T$:
\begin{equation}\label{specific_heat}
  C(T) \sim T^\alpha, \qquad \alpha = \frac2z -1. 
\end{equation}
At large $N$, $\alpha=1+8/(\pi^2 N)$.  The first logarithmic
correction to the specific heat was computed in Ref.~\cite{Vafek}; our
formula~(\ref{specific_heat}) sums up all powers of $\log(T)/N$.

The power-law behavior of the velocity~(\ref{dispersion}) tells us
that at $g\to\infty$ the system is scale invariant with respect to the
scaling transformation~(\ref{scale-z}).  In this regime, it is more
convenient to define the dimensions of the operators with respect to
the scale transformation~(\ref{scale-z}) rather than the relativistic
version~(\ref{scale-1}).  In this new scheme,
\begin{equation}
 [x] = -1, \qquad [t] = -z, \qquad [A_0] = z, \qquad [g^{-2}]=1-z>0.
\end{equation}
The last equation means that the bare kinetic term for $A_0$ is a
relevant perturbation at the strongly coupled fixed point.

\section{Finite Coulomb interaction}
\label{sec:finite-lambda}

\begin{figure}[ht]
  \centerline{\epsfig{file=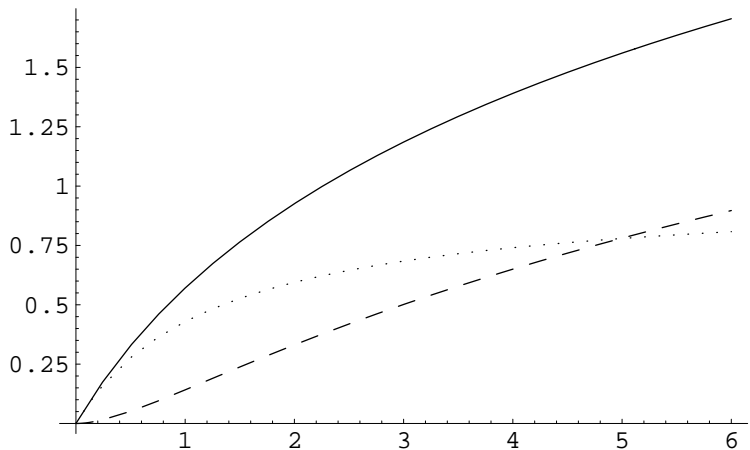,width=0.6\textwidth}}
  \caption[]{The functions $f_1(\lambda)$ (solid line) and 
   $f_0(\lambda)$ (dashed line) and their difference
  $f_1(\lambda)-f_0(\lambda)$ (dotted line).}
  \label{fig:f0f1}
%\vspace{-1ex}
\end{figure}

Let us briefly consider the case of finite $\lambda$.  The loop
integral in Eq.~(\ref{Sigmap}) can be evaluated explicitly (see the Appendix),
%e.g., by expanding 
%the photon propagator in
%geometric series 
%and using the Feynman integration formula (or by any
%\begin{equation}
%  \frac1{A^\alpha B^\beta} = \frac{\Gamma(\alpha+\beta)}
%  {\Gamma(\alpha)\Gamma(\beta)} \int_0^1\!dx\,
%  \frac{x^{\alpha-1}(1-x)^{\beta-1}}{[xA+(1-x)B]^{\alpha+\beta}}\,,
%\end{equation}
%and then resumming the resulting series.  We obtain
\begin{equation}\label{Sigmaf0f1}
  \Sigma(p) = \frac4{N\pi^2}
  [ f_0(\lambda) p_0 \gamma^0 + f_1(\lambda) \p\cdot\vgamma]
  \ln\frac{\Lambda_0}{\Lambda_1}\,,
\end{equation}
where 
\begin{equation}
  \lambda = \frac{g^2 N}{16 v}
\end{equation}
is the parameter measuring the importance of photon self-energy
compared to its bare inverse propagator.  The functions $f_0$ and $f_1$
in Eq.~(\ref{Sigmaf0f1}) are
\begin{equation}\label{f1expl}
  f_1(\lambda) = \left\{ \begin{array}{ll}
    -\displaystyle{\frac{\sqrt{1-\lambda^2}}\lambda}\arccos\lambda 
       -1+ \frac\pi{2\lambda}\,,  & \lambda<1,\stru\\
    \displaystyle{\frac{\sqrt{\lambda^2-1}}\lambda} 
    \ln\left(\lambda+\sqrt{\lambda^2-1}\right)
       -1+ \frac\pi{2\lambda}\,,  \quad & \lambda>1,
  \end{array}\right.
\end{equation}
\begin{equation}\label{f0expl}
  f_0(\lambda) = \left\{ \begin{array}{ll}
    -\displaystyle{\frac{2-\lambda^2}
      {\lambda\sqrt{1-\lambda^2}}}\arccos\lambda 
       -2+\frac\pi\lambda\,, & \lambda<1,\stru\\
    \displaystyle{\frac{\lambda^2-2}{\lambda\sqrt{\lambda^2-1}}} 
    \ln\left(\lambda+\sqrt{\lambda^2-1}\right)
        -2+\frac\pi\lambda\,, \quad & \lambda>1.
  \end{array}\right.
\end{equation}
The asymptotics of the functions $f_0$ and $f_1$ at large and small
$\lambda$ are:
\begin{align}\label{f-asympt}
  f_1(\lambda) &= \left\{ 
    \begin{array}{ll} 
    \displaystyle{\frac\pi4}\lambda
    -\displaystyle{\frac{\lambda^2}3} + O(\lambda^3), & \lambda\ll1, \stru\\
       \ln(2\lambda)-1 + \displaystyle{\frac\pi{2\lambda}}
       +O(\lambda^{-2}), \quad &\lambda\gg1,
    \end{array}\right.\\
  f_0(\lambda) &= \left\{ 
    \begin{array}{ll} 
      \displaystyle{\frac{\lambda^2}3} + O(\lambda^3), & \lambda\ll1, \stru\\
       \ln(2\lambda)-2 + \displaystyle{\frac\pi\lambda}
       +O(\lambda^{-2}), \quad  &\lambda\gg1.
    \end{array}\right.
\end{align}
These functions are related by $f_0(\lambda)=-\lambda
f_1'(\lambda)+f_1(\lambda)$ and are plotted in Fig.~\ref{fig:f0f1},
together with the difference $f_1-f_0$.  The expressions above are
consistent with those quoted in Ref.~\cite{GonzalezGuineaVozmediano}.

%There are two effects on the fermion kinetic term after the
%integration of modes between $\Lambda_1$ and $\Lambda_0$.  The first
%is renormalization of wave function, coming from the correction to
%$\bar\psi\gamma^0\d_0\psi$.  The second effect is the renormalization
%of the fermion velocity, which is due to the fact that the
%$\bar\psi\gamma^i\d_i\psi$ and $\bar\psi\gamma^0\d_0\psi$ receive
%different renormalizations ($f_1(\lambda)\neq f_0(\lambda)$).  One can
%repeat the RG steps many times, integrating out more and more shells
%of momenta with decreasing radii.  

The running of the fermion velocity
is governed by the RG equation
\begin{equation}\label{RG-vlambda}
  p \frac{\d v(p)}{\d p} = -\frac4{\pi^2N} [f_1(\lambda)-f_0(\lambda)]
  v(p) \equiv \gamma_v(\lambda) v(p).
\end{equation}
Since $f_1(\lambda)>f_(\lambda)$ for all $\lambda$ (see
Fig.~\ref{fig:f0f1}) the fermion velocity increases monotonically as
one decreases the momentum scale.  Here, $\gamma_v$ is
the anomalous dimension for the velocity $v$.  As expected, in the limit
of infinite Coulomb coupling it approaches the constant value found above:
\begin{equation}
  \lim_{\lambda\to\infty}\gamma_v(\lambda) = -\frac4{\pi^2 N}\,.
\end{equation}

\section{Finite $N$ and graphene}
\label{sec:graphene}

So far, we have discussed the limit $N\gg1$ where reliable calculations
can be performed.  Let us now discuss finite $N$, keeping in mind that
in graphene $N=2$.  Unfortunately, there is no reliable calculation
tool that works for all values of $N$ and the coupling constant $g$;
therefore, our discussion will be mostly conjectural.

\begin{figure}[ht]
  \centerline{\epsfig{file=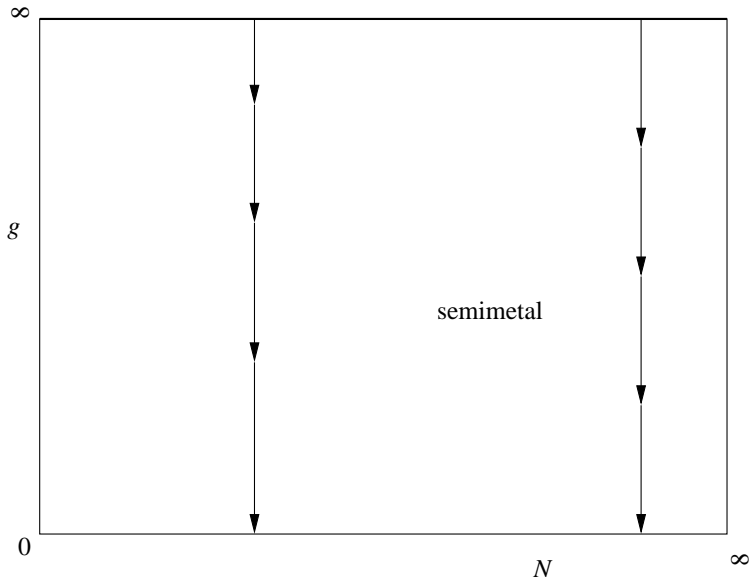,width=0.6\textwidth}}
  \caption[]{The simplest phase diagram.  The line $g=\infty$ is a line of 
  infrared unstable fixed points.  From any finite value of $g$, the system 
  flows to an infrared stable fixed point at $g=0$.}
  \label{fig:phasediag1}
%\vspace{-1ex}
\end{figure}

The simplest possibility is that what we saw at large $N$ remain
qualitatively valid at all $N$: there are two fixed points, an
infrared unstable one at $g=\infty$ and an infrared stable one at
$g=0$.  At $g=\infty$, the system exhibits invariance with respect
to Eq.~(\ref{scale-z}), although the value of $z$ for small $N$ cannot be
computed in a reliable fashion.  This possibility is illustrated in
Fig.~\ref{fig:phasediag1}.  If this is the case, then real graphene
(with $N=2$) is always in the semimetal phase.

\begin{figure}[ht]
  \centerline{\epsfig{file=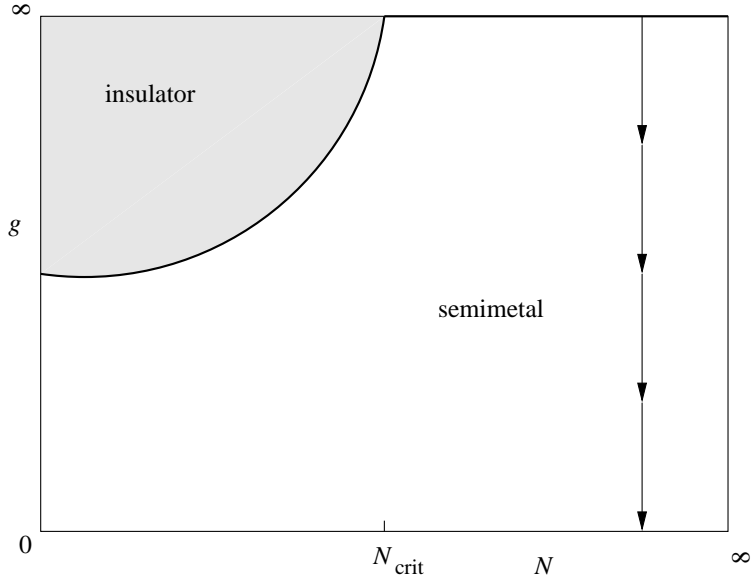,width=0.6\textwidth}}
  \caption[]{A slightly more complicated phase diagram.  For 
  $N>N_{\rm crit}$, the system is always in the semimetal phase.  
  For $N<N_{\rm crit}$, it can be in the semimetal phase [for $g<g_c(N)$]
  or in the insulating phase [for $g>g_c(N)$].}
  \label{fig:phasediag}
%\vspace{-1ex}
\end{figure}

Another possibility is that at sufficiently small $N$, the Coulomb
interaction is strong enough to induce a spontaneous condensation of
particle-hole pairs, creating an excitonic gap which makes the system
insulating. (The alternative possibility of ferromagnetism was 
considered in Refs.~\cite{CastroNeto2,CastroNeto3}.)
This possibility is depicted in Fig.~\ref{fig:phasediag}.
The insulator phase exists when $N<N_{\rm crit}$, but only for
sufficiently strong coupling $g>g_c(N)$.  When $N=N_{\rm crit}$,
$g_c=\infty$, and for $N>N_{\rm crit}$, the insulator phase no longer
exists; the system is in the semimetal phase for all $g$.
Relativistic massless (2+1)D QED is thought to develop a gap when
number of fermion species is below some critical
value~\cite{Appelquist:1988sr}.

If the phase diagram is as in Fig.~\ref{fig:phasediag}, then two
possibilities exist for real graphene.  If $N_{\rm crit}<2$, or if
$N_{\rm crit}>2$ but the bare Coulomb coupling (in vacuum)
$g<g_c(N=2)$, then graphene is a semimetal.  In contrast, if $N_{\rm
crit}>2$ and in vacuum $g>g_c(N=2)$, then freely suspended
graphene~\cite{suspended} is an insulator.  All available experimental
data, on the other hand, are consistent with graphene on a SiO$_2$
substrate being a semimetal.  Therefore, if in vacuum graphene is
insulating, then it undergoes an insulator-semimetal phase transition
as a function of the dielectric constant of the substrate.

The authors of Refs.~\cite{Gorbar:2002iw,Leal:2003sg} solved 
a gap equation with
the screened Coulomb interaction and found $N_{\rm crit}\approx2.55$.
If this is the case, then the system with $N=2$ develops an
excitonic gap at sufficiently large $g$. However, this result is
probably not conclusive as the gap equation in 
Ref.~\cite{Gorbar:2002iw,Leal:2003sg}
is not systematic at small $N$, and also neglects the frequency
dependence in the photon propagator.  Ultimately, the phase diagram of
graphene should (and could) be determined by a direct numerical
simulation of the theory.

In the rest of this section, we shall assume that the phase diagram is
as in Fig.~\ref{fig:phasediag1}, or as in Fig.~\ref{fig:phasediag}
with $N_{\rm crit}<2$.  The system with $N=2$ and infinite Coulomb
coupling should show scale invariance~(\ref{scale-z}) with a critical
exponent $z$ which can be estimated by extrapolating Eq.~(\ref{z}),
\begin{equation}
  z\approx0.8\quad \textrm{for $N=2$}.  
\end{equation}
Note that the $1/N$ correction to $z$ is only $20\%$ at $N=2$, giving
us hope that the $1/N$ expansion works reasonably well even for $N=2$.

The question now is whether in real graphene $\lambda$ is large enough
so that the system is closed to the scale-invariant regime.  If we
take the typical experimental value
$v=10^6~\textrm{m}/\textrm{s}$~\cite{Novoselov-Nature,Kim-Nature} then
we find $\lambda\approx3.4$, which is reasonably large compared to
1.  Therefore, one concludes that graphene is sufficiently close to
the scale-invariant regime.  However, in most experiments the graphene
single layer lies on a substrate, which reduces the effective Coulomb
potential [Eq.~(\ref{g})].  If we take the substrate to be SiO$_2$
with $\varepsilon=4.5$, we find $\lambda\approx1.25$, which is only
marginally larger than 1.

The above quoted value $\lambda\approx3.4$ should be thought of as the
initial condition for the RG equation, valid at the momentum scale of
the order of the inverse lattice size.  As we flow to the infrared, the
system deviates more and more from the strongly coupled fixed point
due to the bare kinetic term which is a relevant perturbation.
However as the initial condition is relatively close to the fixed
point, it will take a large ``RG time'' to go to the weak coupling
regime $\lambda\ll1$.  The graphene thus remains close to the strong
coupling limit for a large momentum range.  The width of this range
can be roughly estimated as
\begin{equation}
  \exp\left(\frac{\pi^2N}4\ln 3.4\right) \sim 10^3,
\end{equation}
where we have used the leading $1/N$ result for the dimension of the
bare Coulomb kinetic term.

An additional information for the effect of finite $\lambda$ can be
obtained from the value of the anomalous dimension for the velocity as given
by Eq.~(\ref{RG-v}) at $\lambda=3.4$:
\begin{equation}\label{gamma_v-finite}
  \gamma_v(3.4) \approx 0.7 \gamma_v(\infty) \approx 0.15.
\end{equation}
It is a 30\% reduction of the anomalous dimension $\gamma_v$.  In
Fig.~\ref{fig:f0f1}, we see that for the realistic $\lambda$ the
different $f_1-f_0$ is already a rather flat function of $\lambda$; an
approximate scaling behavior can be expected.

At the asymptotic infrared end of the RG flow is the trivial fixed
point $\lambda=0$, near which the fermion velocity increases linearly
with the logarithm of the momentum [as followed from Eq.~(\ref{RG-v})
and the small $\lambda$ asymptotics in Eqs.~(\ref{f-asympt})],
\begin{equation}
  -p \frac{\d v(p)}{\d p} = \frac{g^2}{16\pi}\,,
\end{equation}
which translates into an increase of $1.26\times
10^6~\textrm{m}/\textrm{s}$ per decade for graphene in vacuum.
However, this regime is achieved only at extremely low momenta after
the system has passed from the vicinity of the strongly coupled fixed
point to the trivial one.

\section{Conclusion}
\label{sec:conclusion}

In this paper, we consider a model of $N$ massless four-component Dirac
fermions interacting through an instantaneous Coulomb interaction.  We
show that in the limit of infinitely strong coupling, the system shows
a scale-invariant behavior~(\ref{scale-z}) which is characterized by a
dynamic critical exponent $z$.  We obtain the value of $z$ at large
$N$.

We also discuss two possibilities for the phase diagram of the system
at finite $N$.  In one of the possibilities, a part of the phase
diagram is occupied by the insulator phase.  It is very interesting if
in freely suspended graphene the Coulomb interaction is strong enough
to open an excitonic gap.  Provided that such a gap is never opened
for $N=2$, we argue that the Coulomb interaction in vacuum is strong
enough so that freely suspended graphene sheets can be considered as
being in the vicinity of the strong-coupling fixed point for a large
range of momentum.

How can our predictions be tested in experiment?  Assume one could
measure with high precision the quasiparticle dispersion curve in
freely suspended graphene.  If the strong-coupling large-$N$ limit is
any guide, then one should have a slight deviation from the linear
dispersion law, perhaps approximately $\omega\sim p^{0.85}$ [see
Eq.~(\ref{gamma_v-finite})].  For graphene on a substrate, the
deviation from the linear law is smaller.
%If no deviation from the linear
%spectrum is found, then the strong-coupling fixed point discussed in
%this paper would have no relevance to real graphene.
%Instead, one may then
%speculate that graphene is close to a quantum critical point with
%finite charge $g$, where $z=1$~\cite{Herbut-when}.  This is also an
%intereresting possibility from the theoretical point of view.

%In conclusion, we note another interesting system where the results in
%this paper can be of importance: bilayer graphene~\cite{Geim-bilayer}.
%There, the very low-energy electronic excitations have linear
%dispersion relations, but with a Fermi velocity one order of magnitude
%smaller than in the single layer graphene~\cite{McCannFalko}.  This
%means that the intial condition for the RG is very close to infinite
%coupling.  The complication there is that around six (out of eight)
%Dirac points the fermion dispersion is anisotropic.  The
%generalization of the calculation done is this paper to bilayer
%graphene is deferred to future work.

Finally, we hope that this work will motivate numerical simulations of
the model~(\ref{S}), which should help clarify it phase structure.

\acknowledgments
I am indebted to A.~Andreev, D.~Cobden, D.~B.~Kaplan, and S.~Sachdev
for discussions.  This work is supported, in part, by DOE Grant No.\
DE-FG02-00ER41132.

\appendix

\section{Calculation of the beta function at large $N$}

\subsection{Coulomb propagator}

To leading order in the large-$N$ expansion, we must resum all fermion
bubble graphs in the Coulomb propagator.  We shall perform
calculations in the unit system where $v=1$ and restore $v$ in final
formulas when needed.  
\begin{figure}[ht]
  \centerline{\epsfig{file=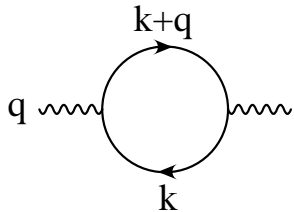,width=0.25\textwidth}}
  \caption[]{The photon polarization diagram.}
  \label{fig:pol}
%\vspace{-1ex}
\end{figure}

The one-loop fermion bubble diagram
(Fig.~\ref{fig:pol}) is
\begin{equation}
  \Pi(q) = N g^2 \int\! \frac{d^3k}{(2\pi)^3}\, \Tr \left(
  \gamma^0 \frac1{\ksl} \gamma^0 \frac1{\ksl+\qsl}\right),
\end{equation}
where $\ksl=\gamma^\mu k_\mu$, $\mu=0,1,2$.  The manipulation of the
Dirac algebra proceeds in a standard fashion.  We use the formulas
\begin{align}
  &\frac1{\ksl} = \frac{\ksl}{k^2}, \\
  &\Tr(\gamma^\mu\gamma^\nu\gamma^\rho\gamma^\sigma) =
  4(\delta^{\mu\nu}\delta^{\rho\sigma} - \delta^{\mu\rho}\delta^{\nu\sigma}
  + \delta^{\mu\sigma}\delta^{\nu\rho})
\end{align}
to find
\begin{equation}
  \Pi(q) = 4Ng^2 \int\! \frac{d^3k}{(2\pi)^3}\,
  \frac{2 k_0 (k_0+q_0) - k\cdot (k+q)}{k^2(k+q)^2} \,.
\end{equation}
We now use Feynman parametrization
\begin{equation}
  \frac 1{AB} = \int_0^1 \!dx\, \frac 1{[xA +(1-x)B]^2} 
\end{equation}
to rewrite
\begin{equation}
  \Pi(q) = 4Ng^2 \int_0^1\!dx\!\int\! \frac{d^3k}{(2\pi)^3}\,
  \frac{2 k_0 (k_0+q_0) - k\cdot (k+q)}{[(1-x)k^2+x(k+q)^2]^2} \,.
\end{equation}
Changing the integration varibale $k\to k-xq$, the denominator in the
integrand becomes an even function of $q$, and one can throw away
terms odd in $k$ in the numerator.  Furthermore due to spherical
symmetry we can replace in the numerator $k_0^2\to k^2/3$.  We find
\begin{equation}
  \Pi(q) = 4Ng^2 \int_0^1\!dx\!\int\! \frac{d^3k}{(2\pi)^3}\,
  \frac{-\tfrac13 k^2 - 2x(1-x) q_0^2 - x(1-x)q^2}{[k^2+x(1-x)q^2]^2}\,.
\end{equation}
We perform integration over $k$ using standard formulas of
dimensional regularization
\begin{align}
   \int\!\frac{d^dk}{(2\pi)^d}\, \frac1{(k^2+\Delta)^n} &= 
  \frac1{(4\pi)^{d/2}}\, \frac{\Gamma(n-\frac d2)}{\Gamma(n)}
  \, \frac 1{\Delta^{n-d/2}}\,,\\
   \int\!\frac{d^dk}{(2\pi)^d}\, \frac{k^2}{(k^2+\Delta)^n} &= 
  \frac1{(4\pi)^{d/2}} \, \frac d2\,  \frac{\Gamma(n-\frac d2-1)}{\Gamma(n)}
  \, \frac 1{\Delta^{n-d/2-1}}
\end{align}
to obtain
\begin{equation}
  \Pi(q) = \frac{g^2N}\pi \frac{|\q|^2}{\sqrt{q^2}} \int\!dx\, 
  [x(1-x)]^{1/2} = \frac{g^2N}8 \frac{\q^2}{\sqrt{q_0^2+v^2|\q|^2}}\,,
\end{equation}
where in the last expression we have restored $v$.  The resummed 
Coulomb propagator is 
\begin{equation}
  D(q) = 
  \left( 2|\q| + \frac{g^2N}8 \frac{|\q|^2}{\sqrt{q^2}}\right)^{-1}
\end{equation}

\subsection{Correction to fermion propagator}

\begin{figure}[ht]
  \centerline{\epsfig{file=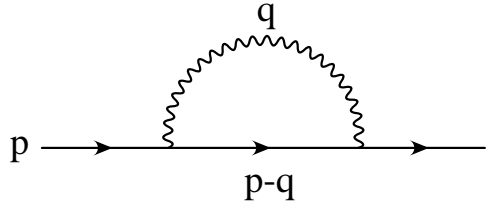,width=0.4\textwidth}}
  \caption[]{The fermion self-energy diagram.}
  \label{fig:self}
%\vspace{-1ex}
\end{figure}

We now compute the $1/N$ correction to the fermion self-energy
(Fig.~\ref{fig:self}):
\begin{equation}
  \Sigma(p) = -g^2 \int\!\frac{d^3q}{(2\pi)^3}\,
  \frac{\gamma^0(\psl-\qsl)\gamma^0}{(p-q)^2} D(q).
\end{equation}

The integral is naively linearly divergent at large $q$, but to
leading order in $q$ the integrand is an odd function of $q$.
Therefore, the integral is only logarithmically divergent and can be
evaluated by expanding in $p\ll q$.  One finds
\begin{equation}
  \Sigma(p) = Z_0 \gamma^0 p_0 + Z_1 \gamma^i p_i,
\end{equation}
where
\begin{align}
  Z_0 &= g^2 \int\!\frac{d^3q}{(2\pi)^3}\, 
    \frac{q_0^2-|\q|^2}{q^4}D(q),\\
  Z_1 &= g^2 \int\!\frac{d^3q}{(2\pi)^3}\, \frac{q_0^2}{q^4}D(q).
\end{align}
Introducing spherical coordinates: $q_0=q\cos\theta$,
$|\q|=q\sin\theta$, one can then write
\begin{align}
  Z_0 &= \frac4{\pi^2N} \, \frac\lambda2\int\limits_0^\pi \!
         \sin\theta\, d\theta\, 
      \frac{\cos^2\theta-\sin^2\theta}{\sin\theta(1+\lambda\sin\theta)}
      \int\!\frac{dq}q\,, \\
  Z_1 &= \frac4{\pi^2N} \, \frac\lambda2\int\limits_0^\pi \!
         \sin\theta\, d\theta\,
      \frac{\cos^2\theta}{\sin\theta(1+\lambda\sin\theta)}
      \int\!\frac{dq}q\,,
\end{align}
where $\lambda=g^2N/16$.  The integral over $dq$ in a spherical shell
$\Lambda_1<q<\Lambda_0$ yields $\ln(\Lambda_0/\Lambda_1)$.  The
angular integral can be taken explicitly and results in
Eqs.~(\ref{Sigmaf0f1}), (\ref{f1expl}), and (\ref{f0expl}).

\end{document}